\documentclass[12pt]{article}

\usepackage{times,amsmath,graphicx}
\usepackage[absolute]{textpos}

\newenvironment{sciabstract}{%
\begin{quote} \bf}
{\end{quote}}

\title{Mounting Evidence for the Violation of Lepton Flavor Universality}

\author
{Andreas Crivellin,$^{1,2,3}$ Martin Hoferichter,$^{4}$\\
\\
\normalsize{$^{1}$ Physik-Institut, Universit\"at Z\"urich, Winterthurerstrasse 190, 8057 Z\"urich, Switzerland}\\
\normalsize{$^{2}$ Paul Scherrer Institut, 5232 Villigen PSI, Switzerland}\\
\normalsize{$^{3}$ CERN Theory Division, 1211 Geneva 23, Switzerland}\\
\normalsize{$^{4}$ Albert Einstein Center for Fundamental Physics, Institute for Theoretical Physics,}\\
\normalsize{University of Bern, Sidlerstrasse 5, 3012 Bern, Switzerland}\\
\\
\normalsize{E-mail: andreas.crivellin@cern.ch, hoferichter@itp.unibe.ch}
}

\date{}

\begin{document} 

\begin{textblock}{10}(8.5,1.5)
	{\small CERN-TH-2021-193, PSI-PR-21-26, ZU-TH 55/21}
\end{textblock}


\baselineskip24pt


\maketitle


\begin{sciabstract}
The Standard Model (SM) of particle physics was finalized in its current form in the mid-1970s, and has been extensively tested and confirmed ever since, with the discovery of the Higgs boson in 2012 being the last missing piece. While no new particles have been directly discovered at the Large Hadron Collider (LHC) at CERN so far, precision observables sensitive to quantum effects of new particles have accumulated intriguing hints for physics beyond the SM. All these anomalies can be interpreted from the point of view of lepton flavor universality, i.e., as hints that electrons, muons, and tau leptons differ much more than predicted by the SM. These tensions can be explained by postulating the existence of new exotic particles. Future measurements will be able to conclusively test this hypothesis, potentially providing long-awaited evidence how the SM needs to be extended at high energies.  
\end{sciabstract}

\section*{Introduction}

The SM has been extensively tested and confirmed within the last half-century, with the discovery of the Higgs boson in 2012 
as its ultimate confirmation as the correct effective theory of particle physics. Nonetheless, it is widely accepted that the SM cannot be the fundamental theory at all energies, as, e.g., it cannot account for neutrino masses, for the dominance of matter over anti-matter in our Universe, and for the existence of dark matter. Therefore, a huge number of extensions of the SM have been proposed ever since its inception, many of which predicted new particles within the reach of high-energy colliders. However, as LHC searches so far have failed to produce direct hints for new particles, indirect searches for physics beyond the SM (BSM) via precision experiments have become an increasingly important avenue. 

In the SM, the building blocks of matter, the fermions, appear in three so-called generations (or flavors), which differ in mass but otherwise behave similarly under the SM interactions, e.g., there are two heavy copies of the electron $e$ (called muon $\mu$ and tau lepton $\tau$), and analogously for up  and down quarks, which make up protons and neutrons. Among the indirect BSM searches, flavor observables, i.e., processes looking for rare transitions between these different  flavors of quarks and leptons, are promising since they are stringently suppressed in the SM and thus very sensitive to BSM physics. In particular, in the SM only the interactions with the Higgs boson can differentiate among the three generations, while all other interactions do not, a property known as lepton flavor universality (LFU). In recent years a pattern of anomalies (deviations from the SM predictions) all pointing towards the violation of LFU has emerged. These observations indicate that electrons, muons, and tau leptons behave more differently than previously expected, suggesting that future tests of LFU will play a key role in elucidating how the SM needs to be extended.  

\section*{The anomalous magnetic moment of the muon
$\boldsymbol{(g-2)_\mu}$}

The $g$-factor of charged leptons was a key prediction of quantum mechanics, which puts its value at exactly $2$. Ever since, increasingly precise measurements of the muon's $g$-factor have served as a sensitive test of the SM, which predicts the ``anomalous'' part $g-2$, and first indications for BSM effects were found in 2006 at Brookhaven. This measurement was recently confirmed by the $g-2$ experiment at Fermilab~\cite{Muong-2:2021ojo}, and the combined result displays a $4.2\sigma$ tension with the SM prediction of the ``Muon $g-2$ Theory Initiative''~\cite{Aoyama:2020ynm}. In particular, the absolute size of the BSM contribution required to reconcile theory and experiment is not small when compared to higher-order SM effects, in such a way that new particles cannot be very heavy to account for this measurement. Since any short-range contribution to $g-2$ scales with the lepton's mass, this observable can be considered a probe of LFU.  

\section*{$\boldsymbol{b\to s\mu^+\mu^-}$}

The class of processes commonly denoted by $b\to s\mu^+\mu^-$ involves, at the fundamental quark level, the transition of a heavy bottom quark ($b$) to a strange quark ($s$) and pair of oppositely charged muons. Among the $b\to s\mu^+\mu^-$ processes, the ratios of decay rates $\Gamma$, $R(K^{(*)}) = \Gamma[B \to K^{(*)}\mu^+\mu^-]/\Gamma[B \to K^{(*)} e^+e^-]$ (the $B$ meson, containing the $b$ quark, decays into a $K$ meson, or its excitation $K^*$, and a lepton pair) are particularly prominent and interesting. These ratios are expected to be approximately one in the SM~\cite{Bobeth:2007dw}, a consequence of LFU, with very small theory uncertainties, while the measured values lie significantly below unity. The most precise measurement by LHCb~\cite{LHCb:2021trn} alone quotes a significance of $3.1\sigma$. 

Further observables that support these hints for the violation of LFU include the angular observable called $P_5'$ in $B \to K^*\mu^+\mu^-$ (measuring correlations among the decay products) and the decay $B_s\to \phi \mu^+\mu^-$ ($\phi$ denotes another meson, composed of $s$ quarks). Altogether, these and other anomalies are consistent with all other available measurements of $b\to s\mu^+\mu^-$ transitions, in such a way that global fits routinely find a preference  compared to the SM hypothesis of $>5\sigma$~\cite{Capdevila:2017bsm,Aebischer:2019mlg}, and even slightly more once the latest updates are included. As these processes are suppressed in the SM, the BSM scale could be quite heavy in this case.

\section*{$\boldsymbol{b\to c\ell\nu}$}

Similarly to $R(K^{(*)})$, the ratios  $R(D^{(*)}) = \Gamma[B \to D^{(*)}\tau\nu] / \Gamma[B \to D^{(*)}\ell\nu]$ (with $\ell=e,\mu$ and $\nu$ a neutrino) show deviations from the SM predictions with a combined significance of about $3\sigma$~\cite{HFLAV:2019otj}, supported by a tension in angular observables~\cite{Bobeth:2021lya}. 
In this case, since the transitions are not suppressed, the BSM effects would have to be quite large, and accordingly the associated scale rather low.

\section*{CAA and $\boldsymbol{q\bar q\to e^+e^-}$}

It has been observed that certain nuclear beta decays (a form of radioactive decays) happen less frequently than expected~\cite{Hardy:2020qwl}. This tension, called the Cabibbo Angle anomaly (CAA), displays a significance around $3\sigma$~\cite{ParticleDataGroup:2020ssz}, and
can again be interpreted as a sign that electrons and muons behave more differently than predicted by the SM~\cite{Crivellin:2020lzu}. Furthermore, the CMS experiment at CERN observed more very high-energetic electrons in proton--proton collisions ($q\bar q\to e^+e^-$) compared to muons than expected~\cite{CMS:2021ctt}, pointing as well towards the violation of LFU.

\begin{figure}[t]
\centering
\includegraphics[width=0.5\linewidth]{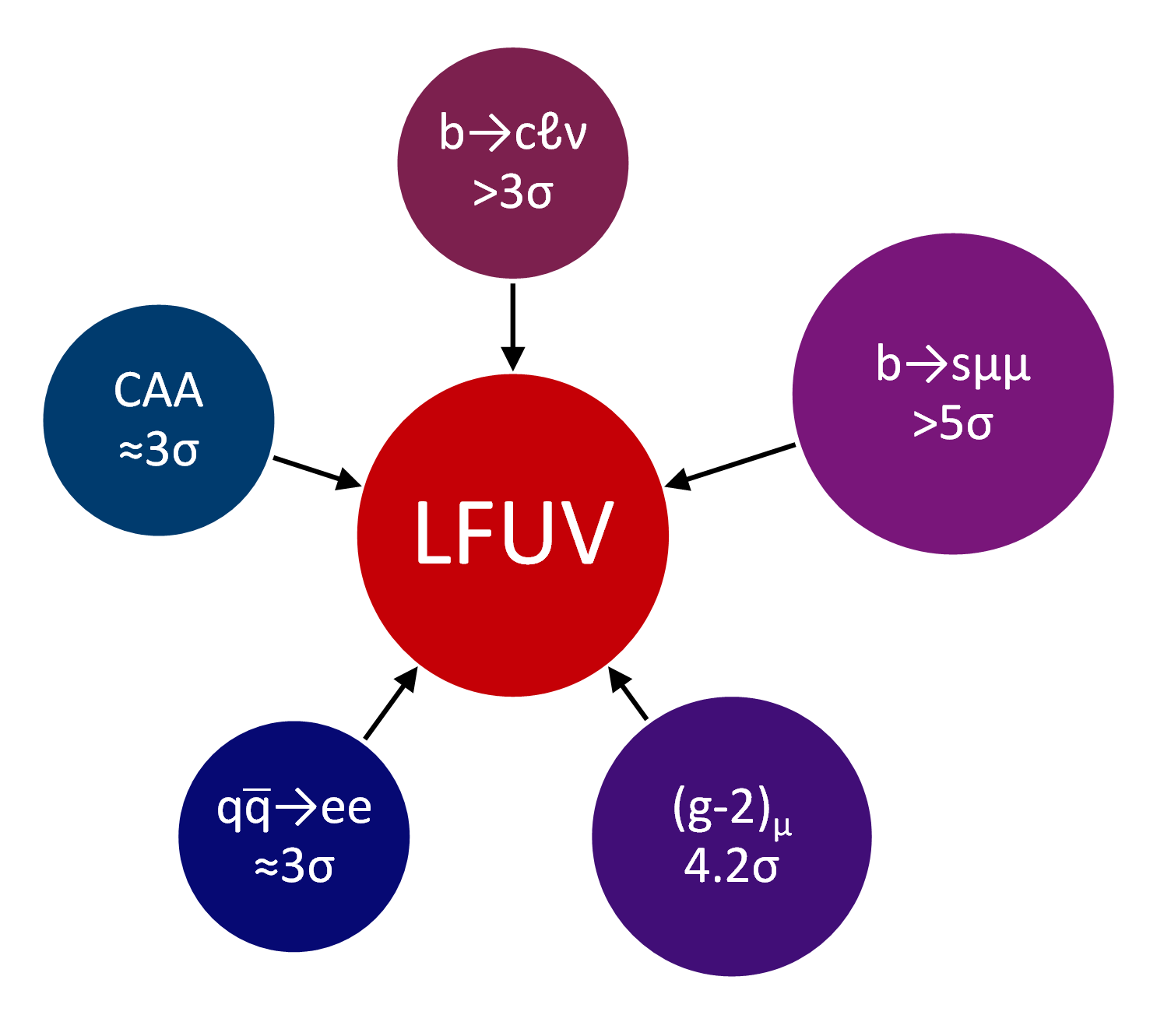}
\caption{Different hints for the violation of LFU, with size of the spheres reflecting the significance of the respective tension.}
\label{fig:LFUV}
\end{figure}

\section*{Explanations and Implications}

With multiple hints for BSM physics pointing towards the violation of LFU, see Fig.~\ref{fig:LFUV}, it is natural to ask how they could be explained in terms of extensions of the SM by new particles and new interactions. 
One promising class of models is known as leptoquarks, hypothetical new particles coupling a quark directly to a lepton, which is forbidden in the SM. Such leptoquarks might be a remnant of grand unified theories, which were devised to unite the different interactions in the SM at high energies.
 
At present, the highest priority is clearly the corroboration of the BSM hints with additional, more precise data, which is in fact ongoing at a number of experiments worldwide. However, even if 
only one of these anomalies were confirmed, this would prove the existence of new particles or interactions at scales that could not only be probed directly at the LHC or a future collider such as the proposed FCC~\cite{FCC:2018byv}, but would also have an impact on other precision observables, allowing for complementary determinations of the new particle's properties. Furthermore, a future electron--positron collider with a sufficiently high energy to produce large quantities of $Z$ bosons, one of the mediators of the weak interactions in the SM, should be able to observe the predicted deviations from the SM expectations in several ways: first, most anomalies, in particular $(g-2)_\mu$, predict effects in $Z$ decays such as $Z\to\mu^+\mu^-$~\cite{Crivellin:2021rbq}. 
Second, the $10^{13}$ $Z$ bosons expected, e.g., for the FCC-ee, would produce an unprecedented number of heavy $b$ quarks and $\tau$ leptons. This would allow for precise tests of the anomalies and correlated observables for which effects are expected, but currently not detectable due to limited statistics. 

Future measurements and improved theory predictions are thus poised to thoroughly scrutinize the currents hints for the violation of LFU. If confirmed, this could provide the long-sought guidance for the construction of the fundamental theory of particle physics, to address the phenomena outside the realm of the current SM, including neutrino masses, dark matter, and the matter over anti-matter asymmetry in our Universe.

\section*{Acknowledgements}

Support by the Swiss National Science Foundation, under Project Nos.\ PP00P21\_76884 (A.C.) and PCEFP2\_181117 (M.H.), is gratefully acknowledged.

\bibliography{scibib}

\bibliographystyle{Science}

\end{document}